\documentclass[preprint,3p,twocolumn]{elsarticle}
\usepackage{graphicx}
\usepackage{amssymb}
\usepackage{lineno}
\journal{Kinematics and Physics of Celestial bodies}
\begin{document}
\begin{frontmatter}
\title{Identification of acoustic gravity waves from satellite measurements}
\author{Yu. O. Klymenko$^*$}
\ead{yurklym@gmail.com}
\cortext[cor1]{Corresponding author.}
\author{A. K. Fedorenko$^1$}
\ead{fedorenkoak@gmail.com}
\author{E. I. Kryuchkov$^2$}
\ead{kryuchkov.ye@gmail.com}
\author{O. K. Cheremnykh$^3$}
\ead{oleg.cheremnykh@gmail.com}
\author{A. D. Voitsekhovska$^4$}
\ead{anna.corvus@gmail.com}
\author{\qquad Yu. O. Selivanov$^5$}
\ead{yuraslv@gmail.com}
\author{I. T. Zhuk$^6$}
\ead{zhukigor@gmail.com}

\address{Space Research Institute, prosp. Akad. Glushkova 40, build. 4/1, 03187, Kyiv, Ukraine}
\begin{abstract}
A method for recognizing the types of linear acoustic gravity waves (AGWs) in the atmosphere from satellite measurements is proposed. It is shown that the polarization relations between fluctuations of wave parameters (velocity, density, temperature, and pressure) for freely propagating waves, as well as evanescent wave modes, differ significantly, which makes it possible to identify different types of atmospheric waves in experimental data. A diagnostic diagram is proposed, with the help of which, from the phase shifts of the observed parameters, it is possible to determine the type of wave, as well as the direction of its movement relative to the vertical. Using phase shifts between fluctuations of the velocity and thermodynamic parameters of the atmosphere, not only the type of the wave, but also its spectral characteristics can be determined. The verification of the proposed method for identification of polar wave disturbances in measurements from the low-orbit satellite Dynamics Explorer 2 was carried out. The verification showed that the polarization relations of AGW in the thermosphere mainly correspond to the gravity branch of acoustic-gravity waves freely propagating in the direction from bottom to top. This conclusion is consistent with other results of AGW observations in the atmosphere and ionosphere by ground-based and satellite methods. No evanescent waves were observed on the considered orbits.
\end{abstract}
\end{frontmatter}

\section{Introduction}
Interest in the study of acoustic-gravity waves (AGWs) in the earth's atmosphere is caused by the fact that the fluxes of energy and momentum carried by them are comparable, and sometimes exceed the energy of other known atmospheric sources \cite{3,4}. AGW significantly affect the development of atmospheric convection, turbulence, the formation of wind flows, which must be taken into account when constructing models of the atmosphere and ionosphere, weather prediction, the study of the propagation of electromagnetic waves, etc.

Wave perturbations of the atmosphere can be described using linearized hydrodynamic equations with respect to the speed of movement of the elementary volume of the medium and its thermodynamic characteristics (density, pressure and temperature) \cite{11,13,15}. From the substitution into these equations of solutions in the form of monochromatic plane waves, the polarization relations are followed, by which you can reveal features of AGW in the atmosphere. The central role in such an analysis plays a dispersion relation, which follows from the condition of the solvability of this linear system of equations. It connects the frequency of AGW with a wave vector. From the dispersion equation, we can conclude about the belonging of observed atmospheric disturbances to acoustic-gravity waves \cite{9,11}.

Polarization relations provide amplitude and phase connections between the oscillations of the AGW speed and thermodynamic fluctuations of density, pressure and temperature, which occur in the atmosphere due to the propagation of the wave. The discrepancy between the theoretical amplitude-phase relations and the experimental data indicates that either these disturbances do not belong to AGWs, or some factors that are significant at observation altitudes were not taken into account in the theoretical calculation of the wave characteristics. This can be a multicomponent chemical composition of the atmosphere \cite{7}, spatially inhomogeneous wind \cite{5}, dissipation or nonlinearity effects. It is also possible a partial coincidence of polarization in some parameters and non-coincidence in others. Such observable features of polarization can be a consequence of limited theoretical ideas about the properties of waves propagating in the atmosphere, and also indicate the direction in which the AGW theory should be developed for better agreement with experiment.

Theoretical analysis of hydrodynamic equations reveals the existence of two main types of AGW in the atmosphere: these are horizontal (evanescent) waves \cite{16} and waves that freely propagate at an angle to the horizontal plane \cite{9,11,15}. The latter are the most common type of waves observed in the earth's atmosphere. Among the evanescent waves also distinguish several known types of waves: the Lamb wave \cite{14}, the Brunt-Väisälä (BV) oscillations \cite{10}, the surface $f$-mode \cite{12}, as well as the recently discovered theoretically $\gamma$-mode \cite{1}. The new approach to the consideration of the evanescent waves was developed in \cite{2}, within the framework of which an infinite set of evanescent solutions were obtained (continuous spectrum).

Simultaneous satellite measurements of different atmospheric parameters (velocity, pressure, density, temperature) give a unique opportunity to identify wave disturbances based on polarization relations.  There are important properties of the propagation of AGW in the atmosphere, which follow from these relations, for example, the difference in acoustic and gravity branches. As is known, with vertical propagation, the energy and phase of acoustic waves propagate in the same direction, and in the case of gravity waves, the phase and energy propagation directions are opposite \cite{9,11}. In horizontal (evanescent) waves, the energy and phase of the waves propagate in the same direction. These and other features are successfully used to interpret atmospheric waves and the improvement of traditional methods of observation of the atmosphere.

In our opinion, among the publications available for today, there is a lack of works on identification of types of atmospheric AGW and the possibility of determining their spectral characteristics according to satellite measurements. In some works on the study of AGW based on satellite data, the properties of free waves were determined using polarization ratios, in particular, the directions of their movement \cite{4,6}. However, in the framework of the general approach, the analysis of the identification of acoustic-gravity waves on the basis of polarization has not yet been carried out.

In this work, we show that different types of AGW have their own characteristic set of phase constraints, which allows to determine the type of wave from observations. In particular, for vertically propagating waves for the component of the vertical velocity $V_z$, in-phase or antiphase disturbances with relative temperature fluctuations $T^\prime/T$ should be observed. The evanescent wave modes should exhibit a phase shift of fluctuations $T^\prime/T, \rho^\prime/\rho$, and $V_x$ with fluctuations of vertical velocity $V_z$ by magnitude of $\pm \pi/2$. For freely (obliquely) propagating waves, fluctuations of different parameters are shifted relative to each other  by an angle, depending on the spectral properties of the wave. Finding such features in experimental data, you can determine the type of wave. We also propose an algorithm for determining the main spectral parameters of AGW according to satellite measurements.

\section{Basic equations}
Consider the movement of the elementary volume of an isothermal atmosphere stratified in the field of gravity. We direct the $z$ axis vertically, and the axis $x$ - along the horizontal component of the particle velocity. In the assumptions of zero viscosity, the absence of wind and the homogeneous chemical composition of the atmosphere, we arrive at a closed system of linearized hydrodynamic equations \cite{13}:
\begin{equation}\label{E:1}
\frac{\partial {{V}_{z}}}{\partial t}+gH\frac{\partial }{\partial z}\left( \frac{{{\rho }'}}{\rho }+\frac{{{T}'}}{T} \right)-g\frac{{{T}'}}{T}=0,
\end{equation}
\begin{equation}\label{E:2}
\frac{\partial {{V}_{x}}}{\partial t}+gH\frac{\partial }{\partial x}\left( \frac{{{\rho }'}}{\rho }+\frac{{{T}'}}{T} \right)=0, \end{equation}
\begin{equation}\label{E:3}
\frac{\partial }{\partial t}\left( \frac{{{\rho }'}}{\rho } \right)+\mathrm{div} \vec{V}-\frac{{{V}_{z}}}{H}=0,
\end{equation}
\begin{equation}\label{E:4}
\frac{\partial }{\partial t}\left( \frac{{{T}'}}{T} \right)+\left( \gamma -1 \right) \mathrm{div} \vec{V}=0.
\end{equation}
The first two equations define the law of the movement of the elementary volume, and the other two are the equation of continuity of the medium and heat equation. The following notation is used: $g$ is the acceleration of gravity, $H=RT/\mu g$ is the atmosphere scale height, $\mu$ is the mean molar mass of the atmosphere, $R$ is a universal gas constant, $T$ is the background temperature, $\rho$ is the density of the atmospheric medium, $V_x$ and $V_z$ are the horizontal and vertical components  of the speed of elementary volume, respectively, $\gamma$ is the ratio of specific heats, $\rho^\prime$ and $T^\prime$ are the perturbed density and temperature, respectively, $\rm{div}\vec{V} = \frac{\partial V_x}{\partial x}+\frac{\partial V_z}{\partial z}$. The relative fluctuation of pressure $P^\prime/P=\rho^\prime/\rho +T^\prime/T$ is from the equation of state of the ideal gas $P = (\rho/\mu)RT$.

We are looking for all solutions (1) - (4) that are proportional to $\exp (i\omega t)$, where $\omega$ is the angular frequency of the wave. Then equations (3) and (4) are reduced to the form:
\begin{equation}\label{E:5}
i\omega \left( \frac{{{\rho }'}}{\rho } \right)={{V}_{z}}\left( \frac{1}{H}-\frac{\mathrm{div}\vec{V}}{{{V}_{z}}} \right),
\end{equation}
\begin{equation}\label{E:6}
i\omega \left( \frac{{{T}'}}{T} \right)=-\left( \gamma -1 \right){{V}_{z}}\left( \frac{\mathrm{div}\vec{V}}{{{V}_{z}}} \right),
\end{equation}
convenient for analyzing the phase shift between the perturbations of the relative fluctuations $\rho^\prime / \rho$ and $T^\prime / T$ with respect to $V_z$. To determine the phase shift between the components of the perturbed velocity, it is convenient to substitute (3) and (4) into equations (1) and (2). As a result, we obtain the equations
\begin{equation}\label{E:7}
{{\omega }^{2}}{{V}_{z}}+g\frac{\partial {{V}_{x}}}{\partial x}+C_{s}^{2}\frac{\partial }{\partial z}\left( \mathrm{div}\vec{V} \right)-\gamma g\mathrm{div}\vec{V}=0,
\end{equation}
\begin{equation}\label{E:8}
{{\omega }^{2}}{{V}_{x}}-g\frac{\partial {{V}_{z}}}{\partial x}+C_{s}^{2}\frac{\partial }{\partial x}\left( \mathrm{div}\vec{V} \right)=0.
\end{equation}
Here $C_s =\sqrt{\gamma g H}$ is the sound speed in the atmosphere.

The condition of solvability of the system of equations (7) and (8) leads to the dispersion equation for AGW.

In the special case of vertically propagating waves, in these equations it is necessary to put ${\partial}/{\partial x}=0$, $V_z\sim \exp (-ik_z z)$, where $k _z$ is the value of the wave vector. Then from equation (6) we obtain
\begin{equation}\label{E:9}
\frac{T'}{T}=(\gamma -1)\frac{{{k}_{z}}}{\omega }{{V}_{z}}.
\end{equation}
Consequently, in a vertically propagating wave, in-phase or antiphase (depending on the sign of $k_z$) is realized for $T^\prime/T$ and $V_z$. Using formula (5), it is easy to see that the oscillations $\rho^\prime/\rho$ and $T^\prime/T$ are shifted relative to each other by a phase angle equal to $\mathrm{arccot} (-Hk_z)$. This allows us to find the wavenumber for vertical AGW, and equation (9) allows us to determine the wave frequency.

\section{Horizontally propagating (evanescent) AGW}

It is convenient to search for wave solutions for evanescent AGWs proportional to a factor $\exp (az-i k_x x)$ having a real value of $a$ \cite{1}. This gives
\[
\frac{\mathrm{div} (\vec{V})}{{{V}_{z}}}=a-i{{k}_{x}}\frac{{{V}_{x}}}{{{V}_{z}}}
\]
The ratio of the velocity components is found from equations (7) and (8):
\[
   \left( {{\omega }^{2}}-k_{x}^{2}C_{s}^{2} \right){{V}_{x}}+i{{k}_{x}}\left( g - aC_{s}^{2} \right){{V}_{z}}=0,
\]
\[
   i{{k}_{x}}\left[ g(\gamma -1) - aC_{s}^{2} \right]{{V}_{x}}
\]
\begin{equation}\label{E:10}
\qquad \qquad \quad  +\left( {{\omega }^{2}}+a^{2}C_{s}^{2}-  \gamma ag \right){{V}_{z}}=0.
\end{equation}
As a result, we get
\begin{equation}\label{E:11}
\frac{{{\rho }'}}{\rho }=i{{V}_{z}}\frac{(a-1/H){{\omega }^{2}}+k_{x}^{2}g(\gamma -1)}{\omega ({{\omega }^{2}}-k_{x}^{2}C_{s}^{2})},
\end{equation}
\begin{equation}\label{E:12}
\frac{{{T}'}}{T}=i{{V}_{z}}(\gamma -1)\frac{a{{\omega }^{2}}-k_{x}^{2}g}{\omega ({{\omega }^{2}}-k_{x}^{2}C_{s}^{2})},
\end{equation}
\begin{equation}\label{E:13}
{{V}_{x}}=i{{V}_{z}}\frac{{{k}_{x}}g(a\gamma H-1)}{{{\omega }^{2}}-k_{x}^{2}C_{s}^{2}}.
\end{equation}
It is seen that the relative fluctuations of $\rho^\prime/\rho$  and $T^\prime/T$ in evanescent waves are in-phase (or antiphase) with respect to each other and are shifted by $\pm \pi / 2$ relative to $V_z$. Also, a phase shift by $\pm \pi / 2$ is between the horizontal and vertical components of the disturbed velocity.

To find the wave parameters $a$, $k_x$ and $\omega$, we add (11) with (12) and take into account equation (13). We also divide (12) by (11). As a result, we obtain the relations
\begin{equation}\label{E:14}
\frac{P'}{P}\equiv \frac{\rho '}{\rho }+\frac{T'}{T}={{V}_{x}}\frac{\Phi }{gH},
\end{equation}
\begin{equation}\label{E:15}
\frac{{{T}'}}{T}=(\gamma -1){{\frac{a{{\Phi }^{2}}-g}{(a-1/H){{\Phi }^{2}}+g(\gamma -1)}}_{{}}}\frac{\rho '}{\rho }.
\end{equation}
Here $\Phi=\omega / k_x$ is the horizontal phase velocity of the wave, which can be found from (14) on the results of observations of $P^\prime/P$ and $V_x$. Substitution of the $\Phi$ value into (15) with the use of the measurement data on the $T^\prime/T$ value allows the $a$ value to be unambiguously determined.

Next, we use the dispersion equation
\[
{{\omega }^{4}}+{{\omega }^{2}}C_{s}^{2}\left( {{a}^{2}}-\frac{a}{H}-k_{x}^{2} \right)+k_{x}^{2}{{g}^{2}}(\gamma -1)=0,
\]
which is obtained from the solvability condition for linear system (10). Rewriting it in the form
\[
k_{x}^{2}=-\frac{{{\Phi }^{2}}C_{s}^{2}a(a-1/H)+{{g}^{2}}(\gamma -1)}{{{\Phi }^{2}}({{\Phi }^{2}}-C_{s}^{2})},
\]
we find the square of the wave vector $k_x$ from the already known values of $a$ and $\Phi$. Then, taking into account the value of $\Phi$, we determine the frequency of the wave, $\omega$. Thus, the observed polarization relation make it possible to find the spectral characteristics of evanescent waves.

\section{Freely propagating AGW}

The solutions of the system of equations (1)-(4) for this type of waves are usually sought in proportion to the spatial factor $\exp \left(z/2H\right) \exp[-i(k_x x + k_z z)]$ [9, 11]. In this case, we obtain the following equation for $\rm{div}\mathnormal{(\vec{V})}$:
\begin{equation}\label{E:16}
\frac{\mathrm{div}(\vec{V})}{{{V}_{z}}}=\frac{1}{2H}-i\left( {{k}_{x}}\frac{{{V}_{x}}}{{{V}_{z}}}+{{k}_{z}} \right),
\end{equation}
as well as polarization relations:
\[
 -\left( {{\omega }^{2}}-C_{s}^{2}k_{x}^{2} \right){{V}_{x}}+C_{s}^{2}{{k}_{x}}\left( {{k}_{z}}-\frac{i\varepsilon}{ H} \right){{V}_{z}}=0,
 \]
\begin{equation}\label{E:17}
   \left( \frac{{{\omega }^{2}}}{{{k}_{x}}} \right)\left( {{k}_{z}}+\frac{i\varepsilon}{H} \right){{V}_{x}}-\left( {{\omega }^{2}}-{{N}^{2}} \right){{V}_{z}}=0.
\end{equation}
Here $\epsilon = 1/\gamma -1/2$, and N is the Brunt-Väisälä frequency, $N^2 = (\gamma - 1) g / (\gamma H)$. Substituting (16) and (17) into equations (5) and (6), we obtain
\[
\frac{\rho '}{\rho }=\frac{{{V}_{z}}}{2H\omega ({{\omega }^{2}}-C_{s}^{2}k_{x}^{2})}
\]
\begin{equation}\label{E:18}
\times \left\{ 2{{\omega }^{2}}({{k}_{z}}H)-i[{{\omega }^{2}}-2\left( 1-{{\gamma }^{-1}} \right)C_{s}^{2}k_{x}^{2}] \right\},
\end{equation}
\[
\frac{T'}{T}=\frac{(\gamma -1){{V}_{z}}}{2H\omega ({{\omega }^{2}}-C_{s}^{2}k_{x}^{2})}
\]
\begin{equation}\label{E:19}
\times \left[ 2{{\omega }^{2}}({{k}_{z}}H)+i({{\omega }^{2}}-2{{\gamma }^{-1}}C_{s}^{2}k_{x}^{2}) \right].
\end{equation}
Hence it follows that the phase shift $\Delta_{\rho , V_z}$ between the oscillations of $\rho^\prime/\rho$ and $V_z$  is determined from the condition
\begin{equation}\label{E:20}
\tan {{\Delta }_{\rho ,{{V}_{z}}}}=-\frac{1-2\left( 1-{{\gamma }^{-1}} \right){{(C_{s}^{{}}{{k}_{x}}/\omega )}^{2}}}{2({{k}_{z}}H)},
\end{equation}
and between the fluctuations $T^\prime/T$ and $V_z$ it is equal to
\begin{equation}\label{E:21}
\tan {{\Delta }_{T,{{V}_{z}}}}=\frac{1-2{{\gamma }^{-1}}{{(C_{s}^{{}}{{k}_{x}}/\omega )}^{2}}}{2({{k}_{z}}H)}.
\end{equation}
The phase shift between the oscillations of the quantities $V_x$ and $V_z$ is found from the first equation in (17):
\[
\tan {{\Delta }_{{{V}_{x}},{{V}_{z}}}}=-\frac{\varepsilon }{{{k}_{z}}H}=
\]
\begin{equation}\label{E:22}
-{{\left( 1-{{\gamma }^{-1}} \right)}_{{}}} \tan {{\Delta }_{T,{{V}_{z}}}}-{{\gamma }^{-1}}\tan{{\Delta }_{\rho ,{{V}_{z}}}}.
\end{equation}
Formulas (20)-(22) demonstrate the existence of phase shifts between fluctuations of different parameters, depending on the spectral properties, which distinguishes them from evanescent waves. Having determined these phase shifts from satellite observations, using formulas (20) and (21), the dimensionless component of the wave vector $k_z H$ and the parameter $(C_s k_x / \omega)^2$ are found:
\begin{equation}\label{E:23}
{{k}_{z}}H={{\frac{1}{2}}_{{}}}\frac{\gamma -2}{{{\left( \gamma -1 \right)}_{{}}} \tan {{\Delta }_{T,{{V}_{z}}}}+\tan {{\Delta }_{\rho ,{{V}_{z}}}}},
\end{equation}
\begin{equation}\label{E:24}
{\left( \frac{{C}_{s} {k}_{x}}{\omega} \right)}^{2}= \frac{\gamma }{2} \frac{\tan \Delta_{T,{V}_{z}}+\tan \Delta _{\rho ,{V}_{z}}}{\left( \gamma -1 \right) \tan \Delta_{T,{V}_{z}}+\tan \Delta_{\rho ,{{V}_{z}}}}.
\end{equation}
To clarify the physical meaning of the $(C_s k_x / \omega)^2$ parameter, we will use the dimensionless dispersion equation for AGW, which follows from (17):
\[
{{\left( \frac{\omega }{N} \right)}^{4}}-\frac{{{\gamma }^{2}}}{\gamma -1}\left[ \frac{1}{4}+{{\left( {{k}_{x}}H \right)}^{2}}+{{\left( {{k}_{z}}H \right)}^{2}} \right]{{\left( \frac{\omega }{N} \right)}^{2}}
\]
\begin{equation}\label{E:25}
+\frac{{{\gamma }^{2}}}{\gamma -1}{{\left( {{k}_{x}}H \right)}^{2}}=0
\end{equation}
Figure 1 shows the regions of existence of the acoustic and gravity branches of the AGW according to the dispersion relation (25). The acoustic region lies above the Lamb line $\omega = C_s k_x$, i.e. here the horizontal phase velocity of waves is greater than the speed of sound $C_s < \omega / k_x$, and the gravity region is below this straight line (the phase velocity of these waves is less than the speed of sound $C_s > \omega / k_x$). Therefore, from the value of $(C_s k_x / \omega)$ found from (24), it is possible to determine whether the observed wave is acoustic or gravity. Equation (23) makes it possible to determine the sign of $k_z$, i.e. the direction of propagation of the wave relative to the vertical. For an acoustic (gravity) wave propagating from below $k_z > 0 ~(k_z <0)$, and from above $k_z < 0 ~(k_z > 0)$.
\begin{figure}[t!]
\centering
\includegraphics[width=\columnwidth]{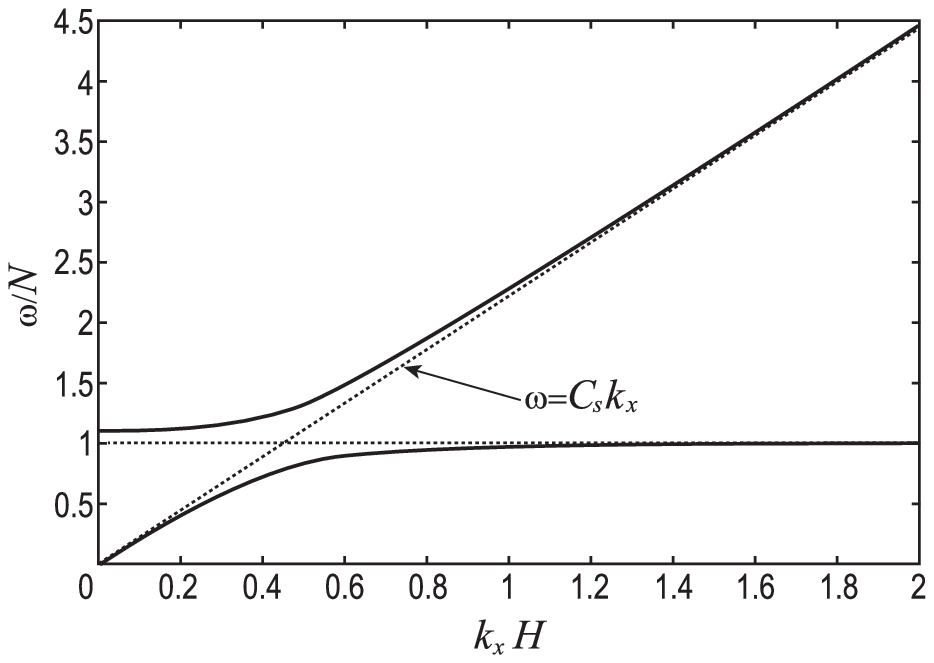}
\caption{Areas of existence of acoustic and gravity waves in an isothermal atmosphere. Dotted lines represent $\omega = C_s k_x$ and $\omega = N$.}
\label{fig:1}
\end{figure}
Further, rewriting the dispersion equation (25) in the form
\[
\left(k_x H \right)^2 \left[ \left( \frac{\omega}{C_s k_x} \right)^2 -1 \right]=
\]
\[
+\left[ \frac{1}{4}-\frac{\gamma -1}{\gamma^2} \left( \frac{C_s k_x}{\omega} \right)^2 \right]+\left( k_z H \right)^2.
\]
and taking into account equations (23) and (24), we can determine the value of the dimensionless component $k_x H$ of the wave vector in terms of the previously introduced angles:
\[
\left(k_x H \right)^2=\frac{\tan \Delta_{T,V_z}+\tan \Delta_{\rho ,V_z}}{\tan \Delta_{T,V_z}-\tan \Delta_{\rho ,V_z}}
\]
\begin{equation}\label{E:26}
\times \frac{(\gamma -1)^2 \left( 1+\tan^2 \Delta_{T,V_z} \right)-\left( 1+\tan^2 \Delta_{\rho ,V_z} \right)}{4 \left[ \left( \gamma -1 \right) \tan \Delta_{T,V_z}+\tan \Delta_{\rho ,V_z} \right]^2}
\end{equation}
As a result, we find the following expressions for the angle $\theta$ between the vector $\vec{k}$ and the horizontal axis $x$:
\[
\tan^2 \theta =\left( \frac{k_z}{k_x} \right)^2=\frac{\tan \Delta_{T,V_z}-\tan \Delta_{\rho ,V_z}}{\tan \Delta_{T,V_z}+\tan \Delta_{\rho ,V_z}}
\]
\begin{equation}\label{E:27}
\times \frac{\left( \gamma -2 \right)^2}{(\gamma -1)^2 \left( 1+\tan^2 \Delta_{T,V_z} \right)-\left( 1+\tan^2 \Delta_{\rho ,V_z} \right)}
\end{equation}
and for the angular frequency of the wave:
\[
\left( \frac{\omega }{N} \right)^2=\frac{\gamma }{2(\gamma -1)}
\]
\[
\times \frac{(\gamma -1)^2 {\tan}^2 \Delta_{T,V_z}+\gamma (\gamma -2) - {\tan}^2 \Delta_{\rho ,V_z}}{\left( \tan \Delta_{T,V_z}-\tan \Delta _{\rho ,V_z} \right)}
\]
\begin{equation}
\times \frac{1}{\left[ \left( \gamma -1 \right) \tan \Delta_{T,V_z}+\tan \Delta_{\rho ,V_z} \right]}.
\end{equation}
To complete the analysis, it is necessary to answer the question, do wave solutions correspond to any values of $\Delta_{T,V_z}$ and $\Delta_{\rho , V_z}$? To get an answer to this question, let us turn to equations (24) and (26), on the left side of which there is a positive value. The requirement that their right-hand sides be positive is in fact equivalent to the condition for the existence of waves. It leads to a system of inequalities, the result of the resolution of which is shown in Fig. 2.
\begin{figure}[h!]
\centering
{\includegraphics[width=\columnwidth]{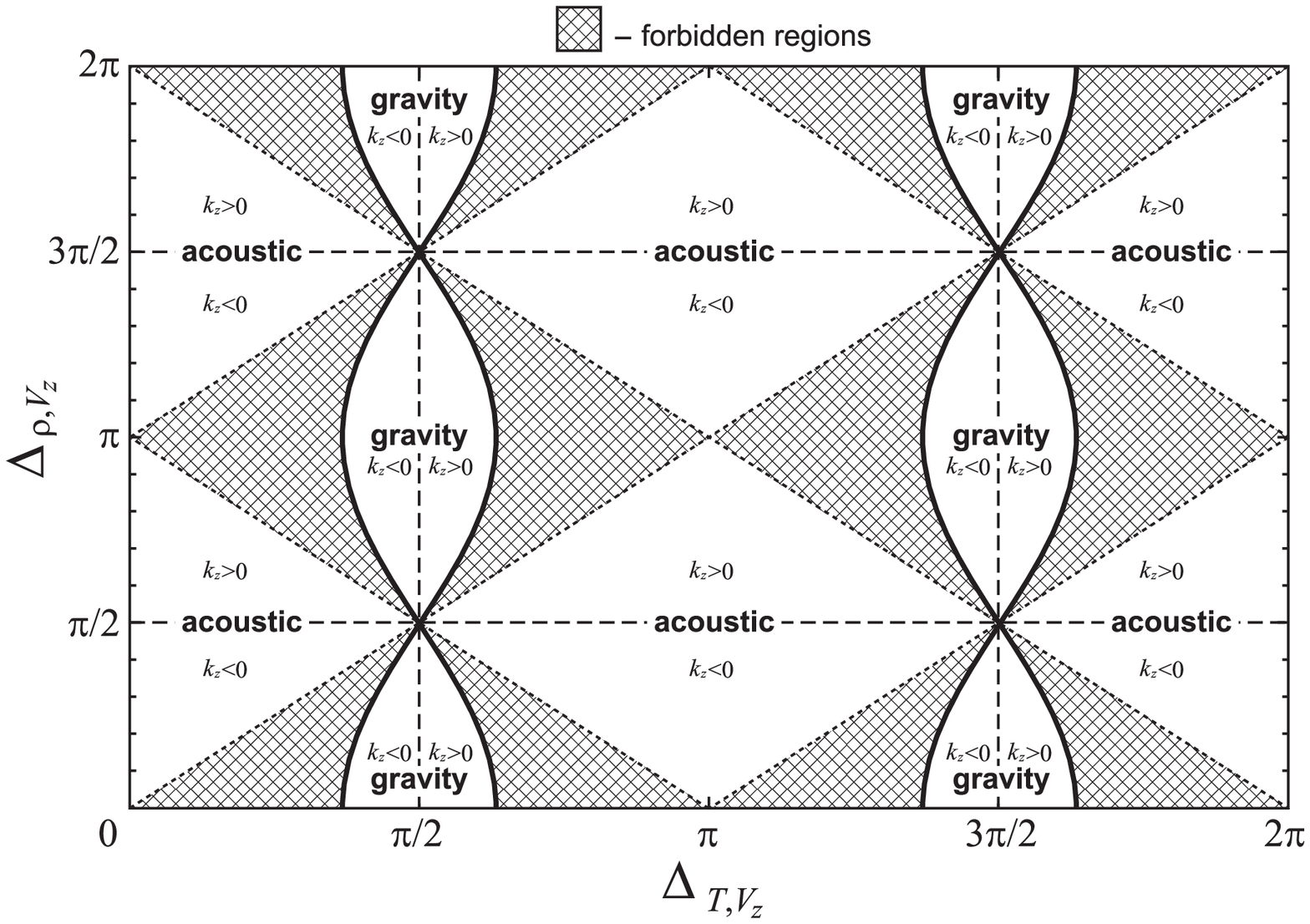}}
\caption{Diagnostic diagram of the regions of existence of gravity and acoustic waves depending on the angles $\Delta_{T,V_z}$ and $\Delta_{\rho , V_z}$ ($0 \le \Delta_{T,V_z} < 2\pi$ and $0 \le \Delta_{\rho , V_z} < 2\pi$).}
\label{fig:2}
\end{figure}
It is seen that in some ranges of phase shifts $\Delta_{T,V_z}$ and $\Delta_{\rho , V_z}$ no waves can exist. These areas in the diagram Fig. 2 are shaded. Here the inequalities $(k_x H)^2<0$ or $(C_s k_x / \omega)^2<0$ are satisfied, i.e. $k_x$ or $\omega$ are purely imaginary values. For all evanescent waves, according to (11) - (13), the phase shifts are $\Delta_{T,V_z}=\pm \pi / 2$, $\Delta_{\rho , V_z}=\pm \pi / 2$, which is hold at the points of intersection of vertical and horizontal dashed lines. Vertical (horizontal) lines divide the gravity (acoustic) areas of internal waves into symmetrical halves with different signs for $k_z$.

\section{Discussion of results and comparison with measurements}

The results established above on the phase shift between the oscillations of physical quantities make it possible to relate the studied atmospheric disturbance with a specific type of AGW. A feature of vertical wave propagation is the in-phase or antiphase of oscillations of temperature $T^\prime/T$ and velocity $V_z$, as well as a nontrivial phase shift between $\rho^\prime/\rho$ and $T^\prime/T$. A distinctive feature of evanescent AGWs is the in-phase or antiphase of oscillations of $\rho^\prime/\rho$ and $T^\prime/T$. Also, evanescent waves should exhibit a phase shift of fluctuations $T^\prime/T$, $\rho^\prime/\rho$, and $V_x$ with a vertical velocity $V_z$ by an amount $\pm \pi / 2$. In a freely propagating inclined wave, the oscillations of the above quantities are spaced relative to each other by a nontrivial angle.

Below, we give an example of diagnosing the AGW type from experimental data and show that the above algorithms allow us to find the spectral characteristics of the wave. Direct measurements from a satellite of various parameters of the atmosphere seem to be the most suitable for testing the proposed identification method.
\begin{figure}[h]
\centering
\parbox{\columnwidth}{\includegraphics [width=0.9\columnwidth] {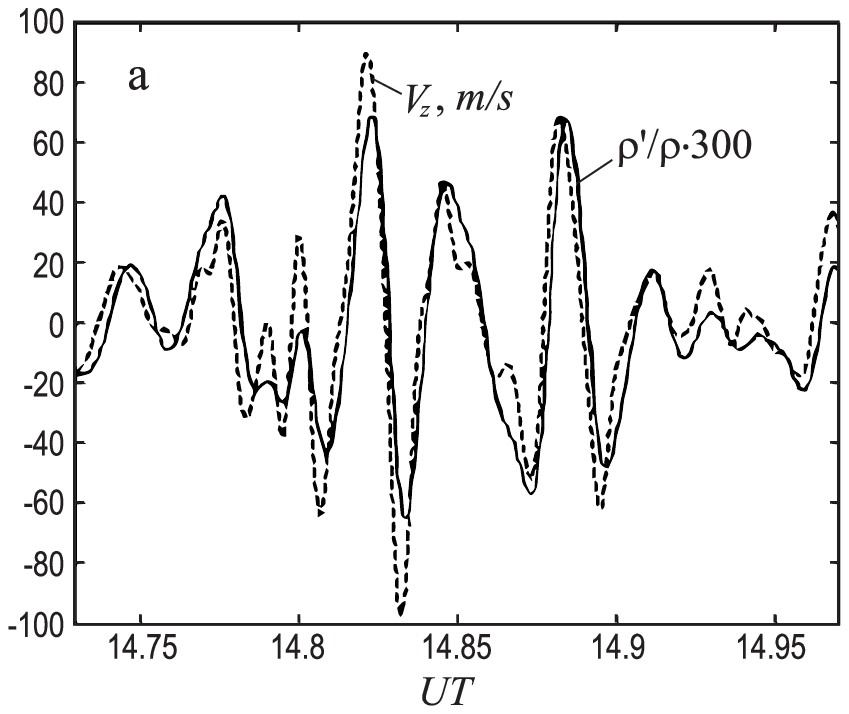}}%
\qquad
\begin{minipage}{\columnwidth}%
\includegraphics [width=0.9\columnwidth] {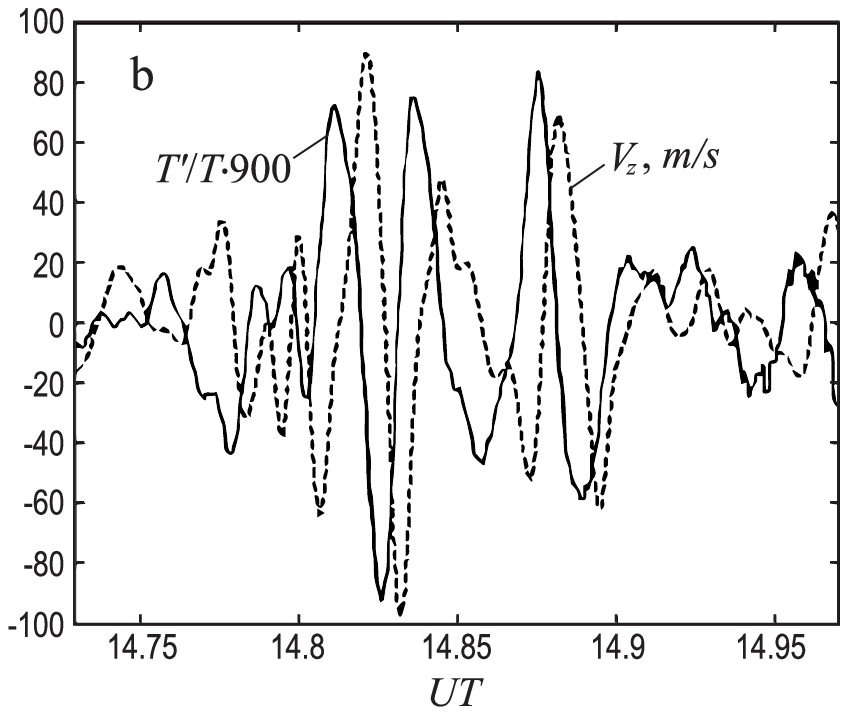}
\end{minipage}%
\caption{Wave fluctuations of density and vertical velocity (a), temperature and vertical velocity (b) on orbit 4820 over the North Polar Cap.}%
\label{fig:3}%
\end{figure}
\begin{figure}[h]
\centering
\parbox{\columnwidth}{\includegraphics [width=0.9\columnwidth] {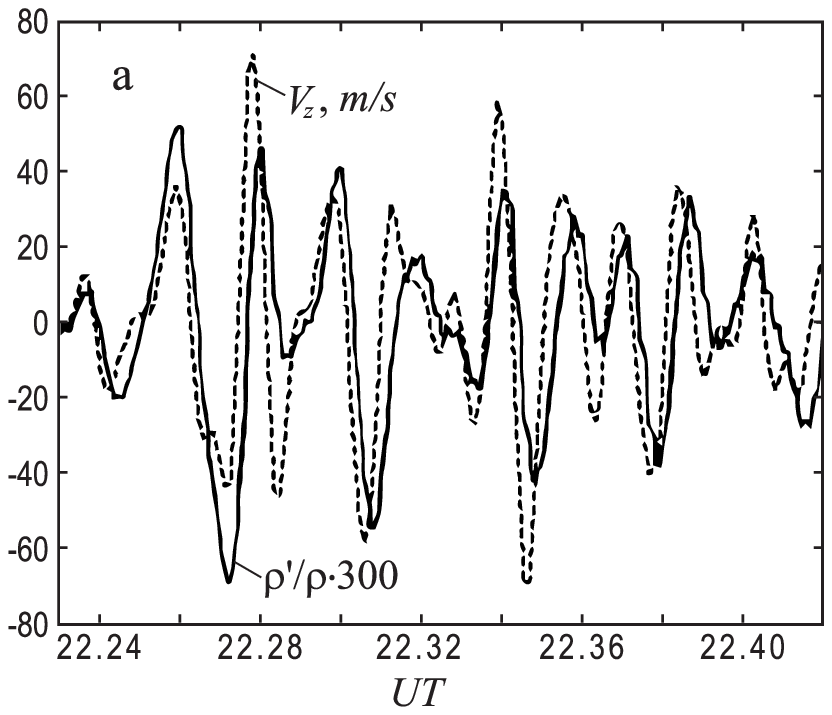}}%
\qquad
\begin{minipage}{\columnwidth}%
\includegraphics [width=0.9\columnwidth] {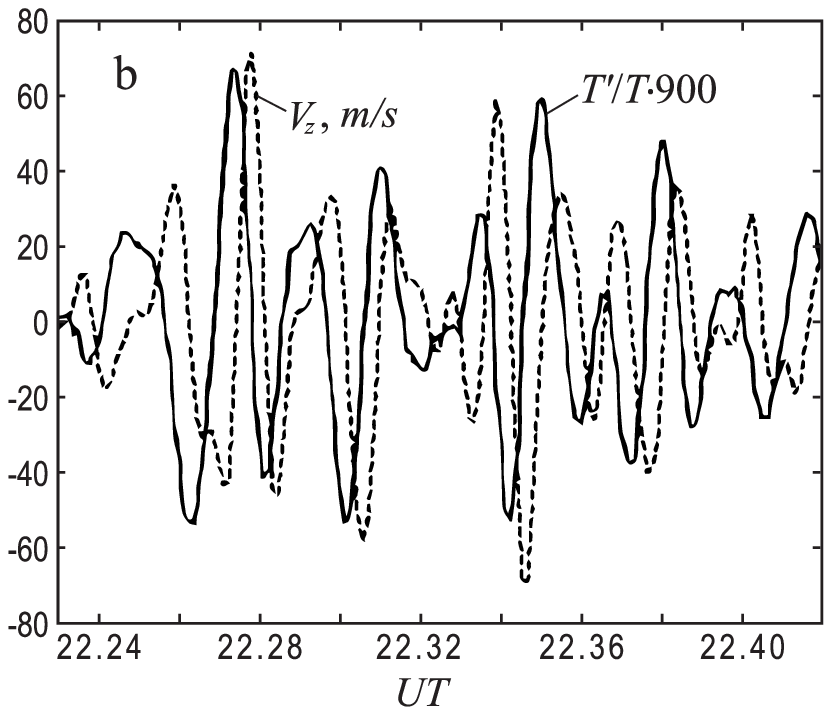}%
\end{minipage}%
\caption{Wave fluctuations of density and vertical velocity (a), temperature and vertical velocity (b) on orbit 8132 over the South Polar Cap.}%
\label{fig:4}
\end{figure}
\begin{figure}[h]
\centering
\parbox{\columnwidth}{\includegraphics [width=0.9\columnwidth] {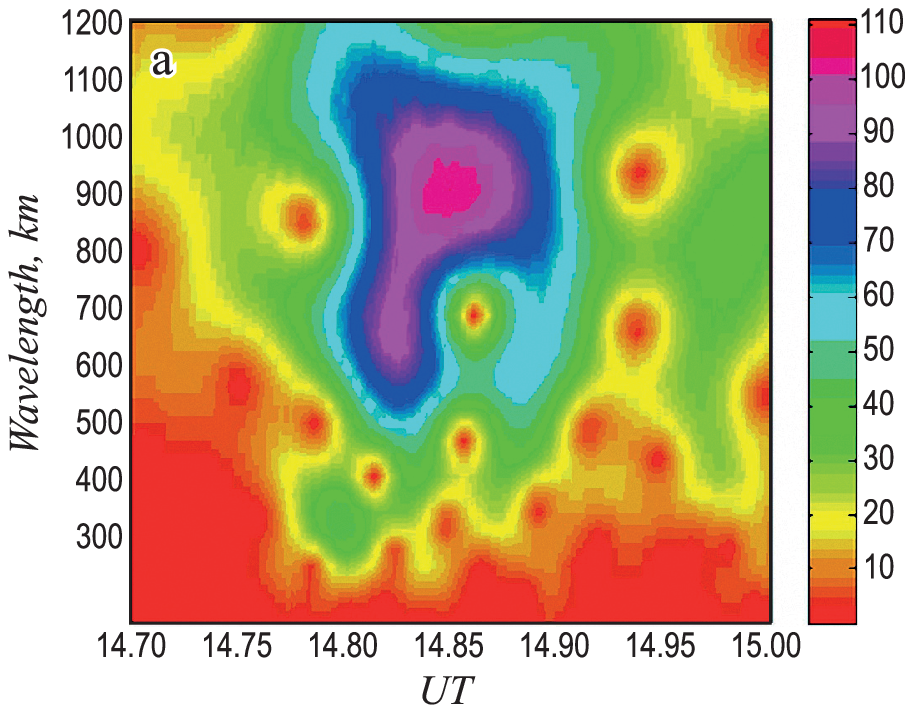}}%
\qquad
\begin{minipage}{\columnwidth}%
\includegraphics [width=0.9\columnwidth] {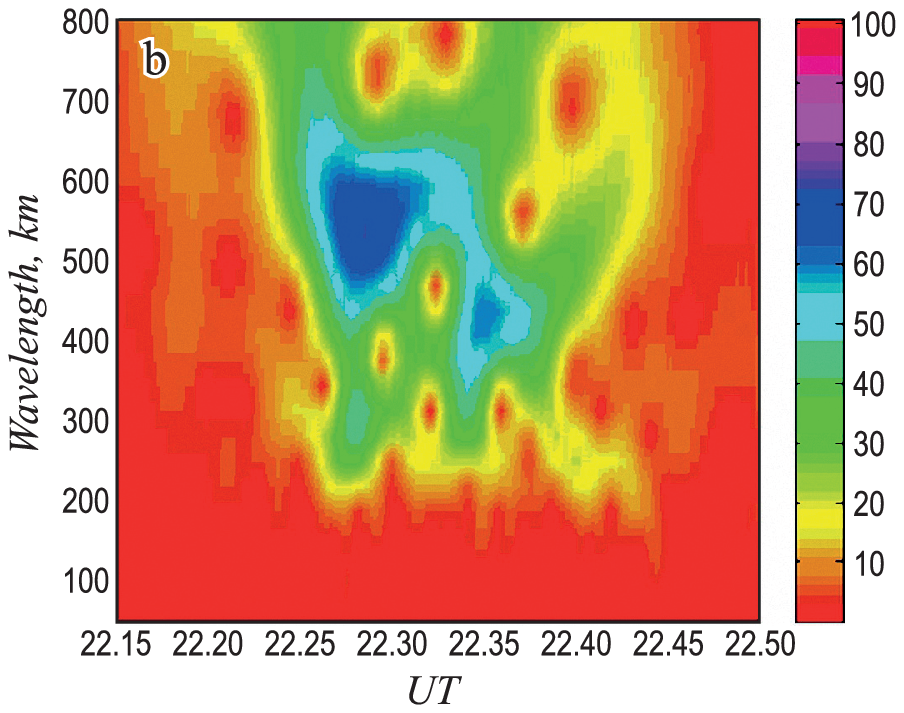}%
\end{minipage}%
\caption{Amplitude wavelet spectra of fluctuations of the vertical velocity component on orbits 4820 (a) and 8132 (b).}%
\label{fig:5}%
\end{figure}
To test the identification method, we chose measurements on the Dynamics Explorer 2 (DE 2) satellite. Measurement data from this satellite are available at ftp:$//$nssdcftp.gsfc.nasa.gov/spacecraft{\_}data. The DE 2 database with the results of direct measurements of atmospheric parameters at ionospheric heights is unique for the study of atmospheric AGW. It is interesting to note that since then no LEO satellites has been launched with such an extensive set of scientific equipment for measuring the parameters of the neutral atmosphere, ionospheric plasma and fields. DE 2 orbital altitude was about 250--1010 km, inclination 89.9$^{\circ}$ (polar orbit), orbital period was about 98 minutes.

The wave frequency measured on the satellite is equal to $\omega^{\prime} = \omega\pm k_{xs}\cdot V_s$, where $\omega$ is the frequency in a rest frame of reference, $k_{xs}$ is the component of the wave vector along the orbit, and $V_s$ is the satellite's velocity. The phase horizontal velocity of the AGW (hundreds of meters per second) is low compared to the speed of the satellite (about 8 km/s). Therefore, the spectrum of wavenumbers $\omega^{\prime} \approx \pm k_{xs}\cdot V_s$ is actually measured on the satellite. Note that the detection of AGW in the atmosphere is complicated by the presence of large-scale changes (trends) associated with the diurnal and geographical variations of atmospheric parameters, changes in the orbital altitude, and other processes. This is most noticeable in measurements of the concentration of neutral particles due to significant variations in the background atmospheric density along the satellite's orbit. To distinguish wave disturbances, these trends must be excluded. Usually the moving average method is used for this \cite{6}.

Over the polar regions, wave disturbances are systematically observed with amplitudes larger than those of middle and low latitudes \cite{6}. These polar waves are characterized by the following features of polarization: 1) oscillations $\rho^\prime/\rho$ and $V_z$ that are close to in-phase; 2) the phase shift between the oscillations $\rho^\prime/\rho$ and $T^\prime/T$ belongs to the interval $(\pi /2) < \Delta_{T,V_z}< \pi$; 3) a small phase shift between the oscillations of the velocity components $V_x$ and $V_z$. The indicated properties of wave fluctuations are typical for the polar regions of the thermosphere \cite{7,8}. Properties 1) and 3) immediately indicate that the observed disturbance is not caused by the evanescent wave. In such a wave, according to (11) - (13), the phase shifts between ($\rho^\prime/\rho$, $V_z$), ($T^\prime/T$, $V_z$), and also ($V_x$, $V_z$) should be $\pm \pi / 2$.

The amplitude-phase features of AGW in the polar regions are illustrated in Fig. 3, 4. On the example of two orbits of the DE 2 satellite, the profiles of oscillations of $\rho^\prime/\rho$ and $V_z$, as well as $T^\prime/T$ and $V_z$, are synchronously shown. The section of orbit 4820 corresponds to a flight over the northern polar cap at an altitude of $\approx$ 320 km, and orbit 8132 - over the southern cap at an altitude of $\approx$ 280 km. From the amplitude wavelet spectra of the $V_z$ component,  shown in Fig. 5, it can be seen that the wave trains in the presented sections are nonmonochromatic and a superposition of oscillations of different scales is observed. Therefore, it is problematic to accurately calculate the phase shift in degrees. Note that although the wavelength along the orbit is not constant, the character of the phase shifts is usually retained over the length of the wave train. In this case, the in-phase, antiphase or phase shift $\pi / 2$ oscillations, when the maximum of one curve corresponds to the zero value of the other curve, can be easily traced in observations.

As seen from Fig. 3 and 4, on orbit 4820 the oscillations of $\rho^\prime/\rho$ and $V_z$ are close to be in-phase, and on orbit 8132 there is a slight phase shift between them. On both orbits 4820 and 8132, the phase shift of $T^\prime/T$ and $V_z$ is close to $\pm \pi / 2$ (zero values on the $V_z$ oscillation profile roughly correspond to the minima and maxima of $T^\prime/T$ oscillations). With such phase shifts, according to the diagram in Fig. 2, the disturbances observed in the polar thermosphere correspond to the gravity branch of the AGW. It can be seen from the diagram that a characteristic feature of the gravity branch is a phase shift about $\pm \pi / 2$ between $T^\prime/T$ and $V_z$.

Since we estimate the phase shift roughly, to find the sign of $k_z$, we turn to the polarization relations (18), (19). The observed oscillations of $\rho^\prime/\rho$ and $V_z$ are close to in-phase, therefore, for the gravity region with $\omega^2 < C_s^2k_x^2$, $k_z < 0$ must be fulfilled. Consequently, the observed AGW propagate from below. In this case, $k_z$ should be small in magnitude so that the phase shift ($T^\prime/T$, $V_z$) is close to $\pi / 2$ according to expression (19). Similar properties of AGW in the polar thermosphere were noted earlier in \cite{5}.

While the type of AGW is quite easy to understand using polarization relations, it is difficult to determine the spectral characteristics of AGW from satellite data. First of all, this is due to the nonmonochromaticity of the observed wave trains. In the general case, the spectral characteristics of AGWs based on phase shifts in different wave parameters can be determined only approximately. In this regard, it is advisable to additionally use the analysis the ratio of the amplitudes of different perturbed quantities, which, in combination with the phase shifts, will make it possible to more accurately estimate the values of the spectral characteristics. At $\Delta_{T,V_z}$ values close to $\pi / 2$, a gravity wave is obtained near the boundary of the free and evanescent regimes with a small $k_z$ value, a period near the Brunt-Väisälä period, and a horizontal wavelength of $\lambda_x = 500-600$ km. Note that even such a rough estimate gives $\lambda_x$ values that are close to the measured values of the $\lambda_{xs}$ projection onto the satellite orbit (see Fig. 5).

\section{Conclusions}

The features of polarization relations between different wave parameters (fluctuations of velocity, density, temperature and pressure) for freely propagating acoustic and gravity waves, as well as evanescent wave modes in an isothermal atmosphere, are analyzed in a linear approximation. For all evanescent waves, the relative fluctuations of temperature $T^\prime/T$ and density $\rho^\prime/\rho$ are observed to be in-phase or antiphase. In addition, the fluctuations of $T^\prime/T$, $\rho^\prime/\rho$, and $V_x$ are phase-shifted with the vertical velocity $V_z$ by the amount $\pm \pi / 2$. For freely propagating waves, fluctuations of different parameters are spaced relative to each other by a certain angle, depending on the spectral properties and belonging to the acoustic or gravity branch of the AGW.

On the basis of the analysis, an algorithm is proposed for identifying the types of acoustic-gravity waves and determining their spectral parameters in the atmosphere from satellite measurements. A diagnostic diagram has been constructed that allows one to determine the type of wave and the direction of its movement along the vertical (up or down) from the observed phase shifts ($\rho^\prime/\rho$, $V_z$) and ($T^\prime/T$, $V_z$). This diagram was used to identify wave disturbances in the polar thermosphere on two orbits of the low-orbit satellite Dynamics Explorer 2. It is shown that the disturbances observed above the polar regions belong to the AGW gravity branch and propagate from below. The spectral properties were determined approximately since the observed wave trains are not monochromatic.

\section*{Acknowledgments}

This work was supported by the National Research Foundation of Ukraine, project 2020.02/0015 "Theoretical and experimental studies of global disturbances of natural and technogenic origin in the Earth-atmosphere-ionosphere system" and partially with the support of the Target Comprehensive Program of the National Academy of Sciences of Ukraine for Scientific Space Research for 2018--2022 years.

\bibliographystyle{elsarticle-num}

\end{document}